\documentclass[aps,prd,showpacs,amsmath,nofootinbib,
superscriptaddress,onecolumn,preprintnumbers]{revtex4}
\usepackage{natbib}
\usepackage{amsmath}
\usepackage{amssymb}
\usepackage{graphicx}
\usepackage{color}
\usepackage{amsmath}
\usepackage{amsthm}
\usepackage{slashed}
\usepackage{graphicx}
\begin{document}

\title{Interplay of CMB Temperature, Space Curvature, and Expansion Rate Parameters}

\author{Meir Shimon}
\affiliation{School of Physics and Astronomy, 
Tel Aviv University, Tel Aviv 69978, Israel}
\email{meirs@tauex.tau.ac.il}
\author{Yoel Rephaeli}
\affiliation{School of Physics and Astronomy, 
Tel Aviv University, Tel Aviv 69978, Israel}
\affiliation{Center for Astrophysics and Space Sciences, University of California,
San Diego, La Jolla, CA, 92093}
\email{yoelr@tauex.tau.ac.il}

\begin{abstract}
The cosmic microwave background (CMB) temperature, $T$, 
surely the most precisely measured cosmological parameter, has been inferred from 
{\it local} measurements of the blackbody spectrum to an exquisite precision of 
1 part in $\sim 4700$. On the other hand, current precision allows inference of 
other basic cosmological parameters at the $\sim 1\%$ level from CMB power spectra, 
galaxy correlation and lensing, luminosity distance measurements of supernovae, as well as 
other cosmological probes. A basic consistency check of the standard cosmological model 
is an independent inference of $T$ at recombination. In this work we first
use the recent Planck data, supplemented 
by either the first year data release of the dark energy survey (DES),  
baryon acoustic oscillations 
(BAO) data, and the Pantheon SNIa catalog, to extract $T$ 
at the $\sim 1\%$ precision level. We then explore correlations between $T$,  
the Hubble parameter, $H_{0}$, and the global spatial curvature parameter, 
$\Omega_{k}$. Our parameter estimation indicates 
that imposing the local constraint from the SH0ES experiment on $H_{0}$ 
results in significant statistical preference for departure at recombination from the locally 
inferred $T$. However, only moderate evidence is found in this analysis 
for tension between local and cosmological 
estimates of $T$, if the local constraint on $H_{0}$ is relaxed. All other dataset 
combinations that include the CMB with either BAO, SNIa, or both, disfavor the addition 
of a new free temperature parameter even in the presence of the local constraint on 
$H_{0}$. Analysis limited to the Planck dataset suggests the temperature 
at recombination was higher than expected at recombination at the $\gtrsim 95\%$ 
confidence level if space is globally flat. An intriguing interpretation of our results is that 
fixing the temperature to its locally inferred value would result in a preference for spatially 
closed universe, if $T(z)$ is assumed to evolve adiabatically and the analysis is based only 
on the Planck dataset.
\end{abstract}

\keywords{}

\maketitle

\section{Introduction}

The cosmic microwave background (CMB) radiation was measured by the COBE/FIRAS 
experiment to have a black body spectral distribution with the exquisitely precise 
temperature $T_{0}=2.72548\pm 0.00057$ at the present epoch [1]. 
Independent inferences of the temperature at redshifts of a few from measurements of 
molecular absorption lines in galaxies at redshifts of a few, and from spectral measurements 
of the Sunyaev-Zeldovich (SZ) effect towards nearby galaxy clusters, though much less 
precise, are still consistent with an adiabatic evolution `history' 
$T(z)=T_{0}(1+z)$, e.g. [2-4] and most recently [5]. 
The scarcity of directly detectable objects at high 
redshifts, and substantial uncertainties in quantifying local astrophysical conditions,  
render these tests rather limited; consequently, it is 
{\it assumed} that adiabatic z-dependence applies (essentially) at all times.

In analyses of datasets that include CMB measurements the value of $T_{0}$ 
is commonly assumed for the temperature [1], but because this physical parameter enters 
in the scalings of key quantities (such as radiation and matter energy densities), and 
essentially determines the timing of the recombination epoch, it is quite important to 
gauge the impact of relaxing this assumption. Clearly, this is of particular interest in 
assessing the precision and possible bias when inferring global parameter values. 
Even though no efficient thermalization mechanism of the CMB past $z\sim 10^{6}$  
is known (or expected), taking the temperature to be a free parameter allows for 
independent determination of its value based largely on CMB features imprinted 
at the recombination era. Assuming that the standard model (SM) reliably describes our universe, 
systematics-free CMB measurements are expected to independently yield a result 
consistent with the locally-inferred value, albeit with a considerably larger statistical 
uncertainty. This test of the SM is well motivated in its own right, and also for 
quantifying any possible deviations from the standard scaling.

For example, it has been shown recently that the (much debated) tension between the 
locally measured value of $H_{0}$, e.g. [6-9], and that inferred from the small 
degree of anisotropy and minute polarization levels of the CMB [10], could 
be explained by deviation from the locally measured value of $T_{0}$, if the adiabatic 
scaling of the temperature with redshift still holds [11, 12]. 
In sharp contrast to distance ladder methods that probe the local 
expansion rate, the CMB provides an inference of $H_{0}$ at recombination, 
$z_{*}\approx 1100$. Earlier studies of the potential impact 
of CMB temperature uncertainties on CMB observables, or on parameter estimation in 
particular, are reported in [13-16].

In this work we explore possible implications that a deduced value of $T_{*}$, 
the value of $T$ at recombination, could have
for the `Hubble tension' and the expected relation to spatial curvature. 
This is based on the realization that a lower than expected $T_{*}$ 
implies earlier decoupling between baryonic matter and radiation, 
which in turn results in a smaller sound horizon at last scattering, i.e. 
smaller angular scale on the sky {\it unless} incoming light rays are 
focused in positively curved space. Thus, a strong 
$T_{*}$-$\Omega_{k}$ correlation would be expected, as has indeed been shown in 
the analysis of [12]. 

On the other hand, assuming flat space, i.e. $\Omega_{k}=0$, it is clear that a lower 
$T_{*}$ implies earlier decoupling and consequently higher inferred $H_{0}$. 
Naturally then, the parameter trio $T - H_{0} - \Omega_{k}$ is the main focus of the 
present work. This relatively simple and flexible picture becomes much more 
constrained when CMB data are supplemented by other large-scale structure (LSS) and 
SN data, although the latter do not explicitly depend on $T$. 
Doing so removes certain parameter degeneracies and results in more stringent 
constraints on key cosmological parameters that correlate with 
$T$, thereby severely limiting the freedom for the latter parameter to stray away 
from the value expected based on the standard adiabatic scaling as is shown below. 

While the idea that $T$ and $H_{0}$ are correlated has been considered 
in [11] \& [12] in the context of the Hubble tension and both analyses 
employed the Planck \& BAO data (the latter was included in the analysis 
in order to break certain parameter degeneracies), these analyses did not use 
the currently available DES 1 yr data \& Pantheon SNIa catalog. In addition, 
model comparison was carried out [11] in a somewhat simplistic 
frequentist approach. No model comparison has been carried out in [12]. In the present work 
we fill in these gaps by including the DES 1 yr \& SNIa datasets, and we do carry 
out a detailed model comparison based on the popular Deviance Information Criterion 
(DIC) -- a compromise between the Bayesian and frequentist approaches -- 
for comparison between the standard and extended cosmological models [17].

The paper is structured as follows. In section II we briefly summarize the status of 
certain known tensions between values of key cosmological parameters.
The extended cosmological model that we consider and assess its statistical 
likelihood is described in section III, followed by an outline in section IV 
of the datasets used in this work and the model comparison criteria we adopt. 
Our main results are described in section V followed by summary in section VI.

\section{Parameter `Tension'}

Currently available high-quality cosmological datasets have 
allowed a sub-percent inference of several key cosmological parameters, 
but the diverse sets of measurements have also  
revealed significant inconsistencies between the deduced values of some of the global 
$\Lambda$CDM parameters. A much-discussed example is the $\gtrsim 5\sigma$ `Hubble tension' between the 
results from cosmologically-based and local distance-ladder-based methods. 
These two fundamentally different approaches for measuring the 
current expansion rate of the universe have been in tension for quite 
some time now, with the Planck data (referred to as P18)
alone favoring a value $H_{0}=67.36\pm 0.54$ km/s/Mpc [10], while a range of 
local measurements seem to indicate systematically higher values, e.g. 
$H_{0}=74.03\pm 1.42$ km/s/Mpc deduced by the SH0ES collaboration by calibrating 
the luminosity of nearby  SNIa (using the period-luminosity relation of Cepheid variable stars [18]). 
Alternative calibration methods infer, e.g. $H_{0}=69.6\pm 1.9$ km/s/Mpc, 
with calibration based on the tip of the red giant branch (TRGB) [19], 
or $H_{0}= 73.3\pm 3.9$ km/s/Mpc by using variable red giant stars for calibration 
[20]. Other local measurements, based on either strong lensing time-delay, e.g. [8, 9] or 
other methods, result in higher-than-Planck values, e.g. [21, 22].   
Compared to local distance-ladder based inferences, CMB anisotropy and polarization 
results are generally considered more reliable due to the fact that the relevant physical 
processes and conditions are well-founded and can be (essentially) fully quantified in the 
linear regime. In addition, some doubts have been raised recently on the validity of 
certain elements of the analysis reported by both the SH0ES collaboration, e.g. [23], 
and the standard treatment using the TRGB approach [24]. Nevertheless, the convergence 
of a range of other approaches based on local measurements towards $H_{0}\gtrsim 70$ 
km/s/Mpc is a rather compelling argument in favor of a claimed `Hubble tension', 
albeit not necessarily as high as is usually claimed. 

In addition, it has been claimed that other 
parameter values are in tension when inferred from different (combinations of) datasets, 
most notably the dimensionless spatial curvature parameter $\Omega_{k}$, 
e.g. [10, 25], and the matter perturbations at a scale of 8 Mpc, $\sigma_{8}$, rescaled to 
$S_{8}\equiv\sigma_{8}/\sqrt{\Omega_{m}}$ in terms of $\Omega_{m}$, 
the current energy density of non-relativistic matter, e.g. [26]. 
Actually, all these are different aspects of the same fundamental tension 
because $H_{0}$, $\Omega_{k}$ and $S_{8}$ are all strongly correlated.

Another anomalous measurement (which is not quantitatively explored in the present work) 
is that of the lensing amplitude of the CMB anisotropy and polarization. It seems that 
there is simply not enough matter in a {\it flat} background universe to explain 
the observed lensing of CMB anisotropy and polarization by the intervening LSS. 
This mismatch is quantified by a dimensionless parameter $A_{lens}$ which should 
be consistent with unity in the SM. 
Observationally, $A_{lens}$ is larger than unity at the $\gtrsim 2\sigma$ confidence 
level. It has already been 
pointed out [25] that $\Omega_{k}$ and $A_{lens}$ are strongly correlated, 
which is indeed expected given the fact that incoming light rays are 
more focused at the background level as $\Omega_{k}$ decreases.

It should be noted that while these tensions are intriguing as they may imply the need for 
new physics, it would 
be statistically more appropriate to quantify the tension between 
datasets (rather than individual parameters) given a multivariate likelihood function 
in terms of, e.g. the `Mahalanobis metric', the `index of inconsistency' [27], or 
the `update difference in mean parameters' [28]. In this spirit, and rather than attempting 
to quantify tension between datasets, we adopt 
the SM comparison approach and look for the best-fit model assuming that the available 
datasets are systematics-free. Nevertheless, the possible resolution or alleviation of 
discordance between certain parameters is a tantalizing possibility that we 
do consider in the present work.

\section{Extended Cosmological Model}

We explore the implications of treating the CMB temperature, whose `local' value was precisely 
measured, as a free parameter. We introduce a new dimensionless parameter $A_{T_{0}}$ -- the 
temperature in units of 2.72548 K, such that $T_{0}$ is replaced 
by $A_{T_{0}}T_{0}$ everywhere in the Boltzmann code CAMB within CosmoMC, 
still maintaining the adiabatic evolution law $T(z)\propto 1+z$. 
The reasoning is clear: $A_{T_{0}}<1$ means that recombination began earlier 
(than with $A_{T_{0}}=1$), implying a smaller acoustic scale at recombination, 
and a higher inferred $H_{0}$. The latter effect can be masked by a spatially 
closed universe ($\Omega_{k}<0$), where incoming light rays are focused and distant 
objects look larger. 
In other words, $A_{T_{0}}$ is expected to correlate with $\Omega_{k}$, as noted above.
These parameters are degenerate with other parameters, which could be 
better constrained in joint analyses of CMB data with other cosmological probes. 

In a vacuum-dominated universe (presumably, the present SM epoch) 
$H_{0}$ is strongly-correlated with $\Omega_{\Lambda}$ and thus similarly 
anti-correlated with $\Omega_{m}$ (because their sum is fixed at unity in a flat spacetime 
with negligible radiation), i.e. a lower 
temperature implies lower values of $\Omega_{m}$ and higher $\Omega_{\Lambda}$. 
In this context we mention that the relative heights of the odd and even acoustic 
peaks in CMB anisotropy, which 
reflect maximum compression and rarefaction of the acoustic plasma waves prior to 
and at last scattering, are regulated by the ratio of $\Omega_{m}/\Omega_{b}$. 
Thus, at least in flat space, any mechanism that boosts the inferred $H_{0}$ value 
(thereby increasing $\Omega_{\Lambda}$ in a DE-dominated, i.e. asymptotically 
de-Sitter, universe) would necessarily lower $\Omega_{m}$, which would in turn 
require a lower $\Omega_{b}$. Consistency of the efficiency of photon-baryon 
interaction prior to decoupling would then require a lower $A_{T_{0}}$, which is 
again consistent with a higher $H_{0}$. However, whereas the physics of the acoustic 
waves is sensitive to the ratio of energy densities 
$\Omega_{rad}/\Omega_{b}$, i.e. to $A_{T_{0}}^{4}/\Omega_{b}$, 
BBN yields sensitively depend on 
the ratio of abundance numbers $\eta_{b}=\Omega_{b}/A_{T_{0}}^{3}$. 
Obviously, it is impossible to keep both ratios fixed exactly to their standard values if 
$A_{T_{0}}$ extrapolated to the present is allowed to depart from $1$. 
However, in practice these values are still free to vary within measurement precision.

Moreover, this relatively simple picture becomes more complicated once $\Omega_{k}$ 
is also treated as a free parameter, as we will see in section V. 
Since the BBN constraints in CosmoMC were calculated assuming the locally 
measured $T_{0}$, we disable the BBN constraints in our simulations. However, we 
do include the helium abundance $Y_{He}$ as a free parameter in our statistical analysis, 
a parameter which could be affected by $A_{T_{0}}$ departure from unity. Ideally, one would couple BBN and 
Boltzmann solver codes for consistency and vary all parameters simultaneously, including 
the baryon density and $A_{T_{0}}$, but this is beyond the scope of the present work.

\section{Datasets and Criteria for Comparative Analysis}

Our baseline model is described by the parameter vector \\ 
$\theta=(\Omega_{b}h^{2}, \Omega_{c}h^{2}, \theta_{MC}, \tau, Y_{He}, A_{s}, n_{s}, 
A_{T_{0}})$, as well as 21 likelihood parameters in case of Planck 2018, 
and 20 additional likelihood parameters in case of DES 1 yr and a few 
more parameters in case of Pantheon datasets, when applicable. 
The cosmological parameters have their standard meanings; 
$\Omega_{b}$, $\Omega_{c}$, are the energy density of baryons 
and cold dark matter in critical density units, respectively; $\theta_{MC}$ 
is the ratio of the acoustic scale at recombination and the horizon scale, 
$\tau$ is the optical depth at reionization, $Y_{He}$ is the helium abundance, 
$A_{s}$ \& $n_{s}$ are the amplitude and tilt of the primordial power spectrum 
of scalar perturbations, respectively. In our analysis $H_{0}$, and the modified mass 
fluctuation $S_{8}=\sigma_{8}\sqrt{\Omega_{m}/0.3}$ (a mass fluctuation measure 
which is optimal for inference by LSS probes such as DES), are derived parameters. 
In our extended model we allow for a non-vanishing spatial curvature 
parameter, $\Omega_{k}$. 

The datasets included here are Planck 2018 temperature anisotropy and polarization as well 
as lensing extraction data, the DES 1yr (cosmic shear, galaxy auto-and cross-correlations), 
BAO (data compilation from BOSS DR12, MGS, and 6DF), 
Pantheon data (catalog of 1048 SNIa in the redshift range $0.01\lesssim z<2.26$), 
and (when applicable) a gaussian prior on $H_{0}$ 
based on local inference by the SH0ES collaboration, all are 
included in the 2019 version of CosmoMC. The Planck likelihood functions employed 
in our analysis include the likelihood function \verb%plikHM_TTTEEE% (the TT, TE and EE 
correlations over the multipole range $30<\ell<2500$), \verb%lowl% \& \verb%lowE% 
(TT and EE auto-correlations over $2\leq\ell\leq 29$), and the CMB lensing likelihood 
function constructed from the 4-point correlation function of the CMB. 
In our analysis we consider five different baseline dataset combinations: 
P18, P18+DES, P18+BAO, P18+Pantheon and P18+BAO+Pantheon. 
We do so with and without the SH0ES prior, as well as with and without curvature 
per each baseline dataset combination. We adopt default CosmoMC flat priors for key cosmological 
parameters which are shown in Table I (along with fiducial values used as initial guess) 
except for $A_{T_{0}}$ whose values are drawn from the range 
[0.5, 1.5]. 

\begin{table}[h]
\begin{tabular} {| c | c | c |}
\hline 
         Parameter & Fiducial & prior \\
\hline
\hline
 {\boldmath$\Omega_b h^2$} & 0.0221 & [0.005, 0.1]\\
 {\boldmath$\Omega_c h^2$} & 0.12 & [0.001, 0.99]\\
 {\boldmath$100\theta_{MC}$} & 1.0411 & [0.5, 10]\\
 {\boldmath$\tau$} & 0.06 & [0.01, 0.8]\\
 {\boldmath$\Omega_{k}$} & 0 & [-0.3, 0.3]\\
 {\boldmath$Y_{He}$} & 0.245 & [0.1 0.5]\\
 {\boldmath$\ln(10^{10}A_{s})$} & 3.1 & [1.61, 3.91]\\
 {\boldmath$n_{s}$} & 0.96 & [0.8, 1.2]\\
 {\boldmath$A_{T_{0}}$} & 1 & [0.5, 1.5]\\ 
\hline
\hline
\end{tabular}
\caption{The basic cosmological parameters, their fiducial values, and flat priors are 
specified. The derived value of $H_{0}$ was constrained to the interval 
[40, 100] km/(s Mpc) .}
\end{table}

Sampling from posterior distributions is done using the fast-slow dragging 
algorithm with a Gelman-Rubin [29] convergence criterion 
$R-1<0.02$ (where R is the scale reduction factor). 
Our analysis of curved models (with a free $A_{T_{0}}$ parameter) using only 
the Planck dataset (with or without the SH0ES prior) did not satisfy 
this convergence criterion; results of this analysis  
are not considered in this work. This is clearly due to the strong correlation of 
$A_{T_{0}}$ with other parameters, especially with $\Omega_{k}$, which is not 
strongly constrained by the Planck data alone. 

Model comparison criteria `penalize' complicated models for additional parameters which 
are not well constrained by the data. The $AIC\equiv\chi_{min}^{2}+2p$ 
(Akaike information criterion) and the $BIC\equiv\chi_{min}^{2}+p\ln N$, where $p$ and $N$ are the free 
model parameters and number of data points used, respectively, represent 
inherently different statistical approaches: Whereas the AIC follows a frequentist 
approach the BIC represents a Bayesian-based decision criterion. In this work we adopt the DIC [17]
\begin{eqnarray}
DIC\equiv 2\overline{\chi^{2}(\theta)}-\chi^{2}(\overline{\theta}),
\end{eqnarray}
where $\theta$ is the vector of free 
model parameters and bars denote averages over the posterior 
distribution $\mathcal{P}(\theta)$. By `Jeffrey’s scale' convention, a model 
characterized by $\Delta\chi^{2}$ (with respect 
to some reference model) $<$1, 1.0-2.5, 2.5-5.0, and $>$5.0 lower than the 
SM (reference model) would be considered as 
inconclusively, weakly/moderately, moderately/strongly, or 
decisively favored [30], respectively. 
While both AIC \& BIC depend only on the peak of the likelihood function, 
the DIC depends on the mean parameter likelihood and on the effective number of parameters 
(i.e. those parameters which are well-constrained by the data).
Another useful diagnostic is the number of parameters actually constrained by the data [17]
\begin{eqnarray}
p_{D}\equiv\overline{\chi^{2}(\theta)}-\chi^{2}(\overline{\theta}).
\end{eqnarray}
This suggests that an extension of a model by $N$ additional parameters is deemed to be 
reasonable if as a result $p_{D}$ increases by $\sim N$.

\section{Results}

The possibility that the $H_{0}$ tension may reflect a more basic $T_{0}$ tension 
has been considered very recently [11, 12]. It was found that for this to be the case, there 
has to be a noticeable departure of the temperature, extracted from a snapshot of the universe at 
$z\approx 1100$, such that $A_{T_{0}}> 1$.
As discussed in the next section (and noted in [12]), this result can be explained by the 
strong $\Omega_{k}-T_{0}$ correlation present in the CMB data that dominates 
over $H_{0}-T_{0}$ anti-correlation expected when (it is assumed that) $\Omega_{k}=0$.
Whereas these analyses were limited only to P18 or P18+BAO data with the SH0ES 
prior, we extend the analysis by including also the DES 1 yr and SNIa data. 
Since $A_{T_{0}}$ is strongly correlated with other cosmological parameters, 
the inclusion of DES and SNIa datasets jointly 
with the P18 and or P18+BAO data severely limits the range over which $A_{T_{0}}$ 
is effectively free to vary. This is akin to the fact that the Planck dataset alone implies that 
$\Omega_{k}<0$ in the SM, but the addition of other probes, most notably BAO, 
`restores' $\Omega_{k}=0$. 

In this work we compare the fit of four different cosmological models to various combinations 
of the datasets. In addition to the flat SM, we refer to the SM with $\Omega_{k}\neq 0$ as `SM+K'; 
the SM with $A_{T_{0}}$ as `SM+T', and to the extension of SM+T model to 
allow for curved space as `SM+K+T'. In Table II we show results obtained 
from the joint analyses of the Planck dataset with the SH0ES prior in the 
SM and SM+T models. We focus on this particular 
dataset combination because it results in the most significant differences between results for 
the SM and SM+T models. Most of the key 
parameters are biased by $\sim 3$ standard deviations in the SM as compared with the 
SM+T model. We define a dimensionless bias for a given parameter $\alpha$ between 
two models `a' and `b' as 
$\delta_{\alpha_{ab}}\equiv (\alpha_{b}-\alpha_{a})/\sqrt{\sigma_{\alpha,a}^{2}+\sigma_{\alpha,b}^{2}}$, 
where $\alpha_{i}$ and $\sigma_{\alpha,i}$ are the values
of $\alpha$ and its uncertainty, respectively, inferred in the i'th ($a$ or $b$) model. 
In case of asymmetric posterior distributions we use 
the largest error of the two sides of the distribution in the denominator 
of $\delta_{\alpha_{ab}}$. In addition, results for the SM seem to 
underestimate the uncertainties in cosmological parameters when ignoring 
their correlations with $A_{T_{0}}$. 
It is clear from the values shown in the rightmost column of the table that the SM 
underestimates the uncertainties in many of the parameters by a factor $\sim 3-7$, 
and up to $\sim 46$ in the extreme case of $\theta_{MC}$. While thawing 
fixed constants always results in increased errors, correlations, and bias, we note 
that if indeed the temperature evolves non-adiabatically with redshift in a fashion that 
is theoretically unknown, then these levels of bias and precision degradation may constitute 
the actual (rather than nominal) precision available at present.

\begin{table}
\begin{tabular} {| c | c | c | c | c |}
\hline 
         & Standard Model   & Standard Model + $T$& &\\
\hline    
         Parameter & 68\% limits & 68\% limits & Bias & Degradation\\
\hline
\hline
 {\boldmath$\Omega_b h^2$} & $0.02259\pm 0.00019$ & $0.0182\pm 0.0012$& $-3.6$ & $6.3$ \\
 {\boldmath$\Omega_c h^2$} & $0.1182\pm 0.0011$ & $0.0980\pm 0.0057$& $-3.5$& $5.2$\\
 {\boldmath$100\theta_{MC}$} & $1.04135\pm 0.00052$ & $1.115_{-0.024}^{+0.021}$& $3.1$& $46.2$\\
 {\boldmath$\tau$} & $0.0604_{-0.0082}^{+0.0070}$ & $0.0562\pm 0.0078$ & $-0.4$ & $1$\\
 {\boldmath$Y_{He}$} & $0.251\pm 0.012$ & $0.239\pm 0.013$ & $-0.7$ & $1.1$\\
 {\boldmath$\ln(10^{10}A_{s})$} & $3.055_{-0.017}^{+0.014}$ & $3.047\pm 0.016$ & $-0.3$ & $1$\\
 {\boldmath$n_{s}$} & $0.9720\pm 0.0067$ & $0.9635\pm 0.0072$ & $-0.9$ & $1.1$\\
 {\boldmath$H_{0}$} & $68.34\pm 0.57$ & $72.5\pm 1.3$ & $2.9$ & $2.3$\\
 {\boldmath$\Omega_{\Lambda}$} & $0.6972_{-0.0065}^{+0.0072}$ & $0.777_{-0.020}^{+0.022}$ & $3.5$ & $3.1$\\
 {\boldmath$\Omega_{m}$} & $0.3028\pm 0.0069$ & $0.223_{-0.022}^{+0.020}$ & $-3.5$ & $3.2$\\
 {\boldmath$\Omega_{m}h^{2}$} & $0.1414\pm 0.0010$ & $0.1169 \pm 0.0069$ & $-3.5$ & $6.9$\\
 {\boldmath$\sigma_{8}$} & $0.8107\pm 0.0070$ & $0.880\pm 0.023$ & $2.9$ & $3.3$\\ 
 {\boldmath$S_{8}$} & $0.814\pm 0.012$ & $0.758\pm 0.020$ & $-2.4$ & $1.7$\\
 {\boldmath$A_{T_{0}}$} & $1.0\pm 0.000209$ & $0.934\pm 0.020$ & $-3.3$ & $95.6$\\ 
\hline
\hline
\end{tabular}
\caption{Marginalized average values of 
key cosmological parameters along with their $68\%$ confidence regions are shown for 
the joint P18+SH0ES analysis applied to the standard cosmological model (second column) 
and its extension with a free $A_{T_{0}}$ parameter (third column) assuming 
$\Omega_{k}=0$. The fourth column shows the relative bias (in $1\sigma$ units) 
between the standard and extended model. For most of the 
parameters the absolute value of the bias is non-negligible $\sim \pm 3$. Since $H_{0}$ and $T$ 
are strongly anti-correlated, a lower $A_{T_{0}}$ implies a larger $H_{0}$, 
the latter ($H_{0}=72.5\pm 1.3$) is now well within the $1-\sigma$ uncertainty of the locally inferred value 
from the SH0ES: $H_{0}=74.03\pm 1.43$. The `degradation' in the precision inference introduced by the addition of $A_{T_{0}}$ 
is shown in the right-most column; it is clear that parameters which are significantly biased 
also suffer from significant degradation commensurate with the interpretation that they are 
most affected by allowing $T_{0}$ to depart from its FIRAS-deduced value. 
The value of $A_{T_{0}}$ quoted in the second column is adopted from [1].}
\end{table}

Results for the four models with various Planck, DES, BAO, SNIa dataset combinations, 
with and without the SH0ES prior, are listed in Table III. 
The first five lines in Table III correspond to datasets that include the SH0ES prior. 
From the DIC values it is clear that both P18+SH0ES and P18+DES+SH0ES 
decisively favor the SM+T over the SM, and `benefit' mainly from 
adding $A_{T_{0}}$ as a free parameter 
(relatively) less so from allowing for a non-flat geometry. The lower DIC 
obtained for the SM+T model with the P18+DES+SH0ES data combination 
results when $A_{T_{0}}<1$, which is due to the 
larger $H_{0}$ by virtue of the anti-correlation of these two parameters.
Specifically, with the P18+SH0ES data we obtain 
$A_{T_{0}}=0.934\pm 0.020$ \& $0.908\pm 0.047$ at the 68\% C.L., 
assuming the SM+T \& SM+K+T, respectively. These correspond to departure from the 
canonical value $A_{T_{0}}=1$ at the $\gtrsim 99.9\%$ \& $\lesssim 95\%$ C.L, respectively. 
For the P18+DES+SH0ES our analysis yields $A_{T_{0}}=0.950\pm 0.013$ \& $0.956\pm 0.025$ at 
the 68\% C.L in the SM+T \& SM+K+T cases, respectively, with similar C.L for departure 
from the canonical value as deduced for the P18+SH0ES combination.

\begin{table}
\begin{tabular} {|c | c | c|c|c|}
\hline 
\hline
Datasets &  $DIC_{SM}$  &$DIC_{SM+T}$ & $DIC_{SM+K}$  & $DIC_{SM+K+T}$\\
\hline
\hline
P18+SH0ES   & 2828.80 & 2818.25 & 2821.87 & -- \\
\hline
         &         & -10.55  & -6.93   & -- \\
\hline
\hline
P18+DES+SH0ES   & 3364.84 & 3352.08 & 3355.07& 3353.05\\
\hline
         &         & -12.76  &  -9.77 & -11.79\\
\hline
\hline
P18+BAO+SH0ES   & 2833.91 & 2834.05 & 2832.06 & 2831.28\\
\hline
             &         & 0.14    & -1.85  & -2.63\\
\hline
\hline             
P18+SN+SH0ES   & 3863.59 & 3858.92 & 3857.52& 3858.49\\
\hline
            &         & -4.67   &  -6.07 & -5.1   \\
\hline
\hline
P18+BAO+SN+SH0ES & 3869.17  & 3869.04 & 3866.86& 3866.94\\
              &          & -0.13   & -2.31  & -2.23\\
\hline
\hline
P18   & 2809.23 & 2806.65 & 2807.72 & -- \\
\hline
         &         & -2.58  & -1.51   & -- \\
\hline
\hline
P18+DES& 3349.71 & 3347.56 & 3349.66& 3344.72\\
\hline
      &         & -2.15   & -0.05  & -4.99\\
\hline
\hline
P18+BAO & 2816.21 & 2816.38 & 2816.87& 2817.17\\
\hline
       &         & 0.17 & 0.66      & 0.96 \\
\hline
\hline
P18+SN & 3844.87 & 3845.63 & 3845.57& 3845.07\\
\hline
      &         & 0.76    & 0.70   & 0.20\\
\hline
\hline
P18+BAO+SN & 3850.74 & 3851.23 & 3851.82& 3851.71\\
\hline
          &         & 0.49    & 1.08   & 0.97\\
\hline
\hline
\end{tabular}
\caption{Model comparison between the SM, SM+T, SM+K, and SM+K+T. 
For each data combination we calculate the DIC for each of the four models, and three 
values of $\Delta DIC$ when compared with 
the SM. We emphasize that while nested models always have lower $\chi^{2}$ values, 
their DIC does not have to be lower than that of the extended model due to the `penalty' 
for additional poorly-constrained parameters. The latter case is best illustrated by the 
$\Delta DIC>0$ values of either P18+BAO, P18+SN or P18+BAO+SN for which none of 
the extended models considered in our work is justified.}
\end{table}

Comparison with LSS probes is especially important as it provides an important 
cross-check. As is well-known, DES yr 1 analysis [26] resulted in 
$S_{8}\equiv\sigma_{8}(\Omega_{m}/0.3)^{0.5}= 0.773_{-0.020}^{+0.026}$ 
and $\Omega_{m}=0.267_{-0.017}^{+0.030}$, with $S_{8}$ 
systematically lower than inferred for the SM with the P18 data.
In the SM, $A_{T_{0}}$ is fixed to unity and so $H_{0}$ cannot increase 
appreciably, even with the inclusion of 
the SH0ES prior (due to its strong anti-correlation with $A_{T_{0}}$) 
and consequently $S_{8}$ cannot correspondingly decrease.
We emphasize that compatibility of Planck with LSS probes, 
such as DES, in the $H_{0}-S_{8}$ plane with the SH0ES 
prior is not guaranteed, and is thus a non-trivial test for any extension 
of the SM purported to address the `Hubble tension'. For example, it has 
been noted recently [31] that `early dark energy' models, e.g. [32], 
while seem to relieve the Hubble tension, bring the CMB- and LSS-derived mass 
clustering parameter $S_{8}$ to a more significant tension level than already exists 
in the SM. To get a sense of the compatibility of the Planck+DES-based inference 
of $S_{8}$ in the models explored in this work we mention that when the P18+SH0ES 
combination is considered $S_{8}=0.814\pm 0.012$, $0.797\pm 0.013$ \& 
$0.758\pm 0.020$ in case of the SM, SM+K \& SM+T, respectively. 
For comparison, when P18+DES+SH0ES is considered we obtain 
$S_{8}=0.8013\pm 0.0099$, $0.790\pm 0.010$, $0.776\pm 0.012$ \& $0.776\pm 0.013$ in 
case of the SM, SM+K, SM+T \& SM+K+T, respectively. 
The lower values for $S_{8}$ obtained in the SM+T \& SM+K+T models for P18+SH0ES 
(i.e. even when the DES is not included in the analysis) bring 
this dataset combination to a better agreement with the DES collaboration results reported 
in [26] by virtue of the strong $A_{T_{0}}-S_{8}$ correlation: A 
higher $H_{0}$ implies lower $A_{T_{0}}$ and 
$S_{8}$ as is clearly demonstrated in Figures 1 \& 2.
In Figure 1 we show the 1- and 2-$\sigma$ confidence contours 
of $\Omega_{b}h^{2}$, $A_{T_{0}}$, $H_{0}$ \& $S_{8}$, along with their posterior distributions 
for the SM+T model (with and without the SH0ES prior and for all five 
data set combinations considered in this work). 
Similarly, shown in Figure 2 are the results for the SM+K+T model, and in Figure 3 
for both the SM and SM+T models based on the Planck dataset with the SH0ES prior.

\begin{figure}[h]
\begin{center}
\leavevmode
\includegraphics[width=0.7\textwidth]{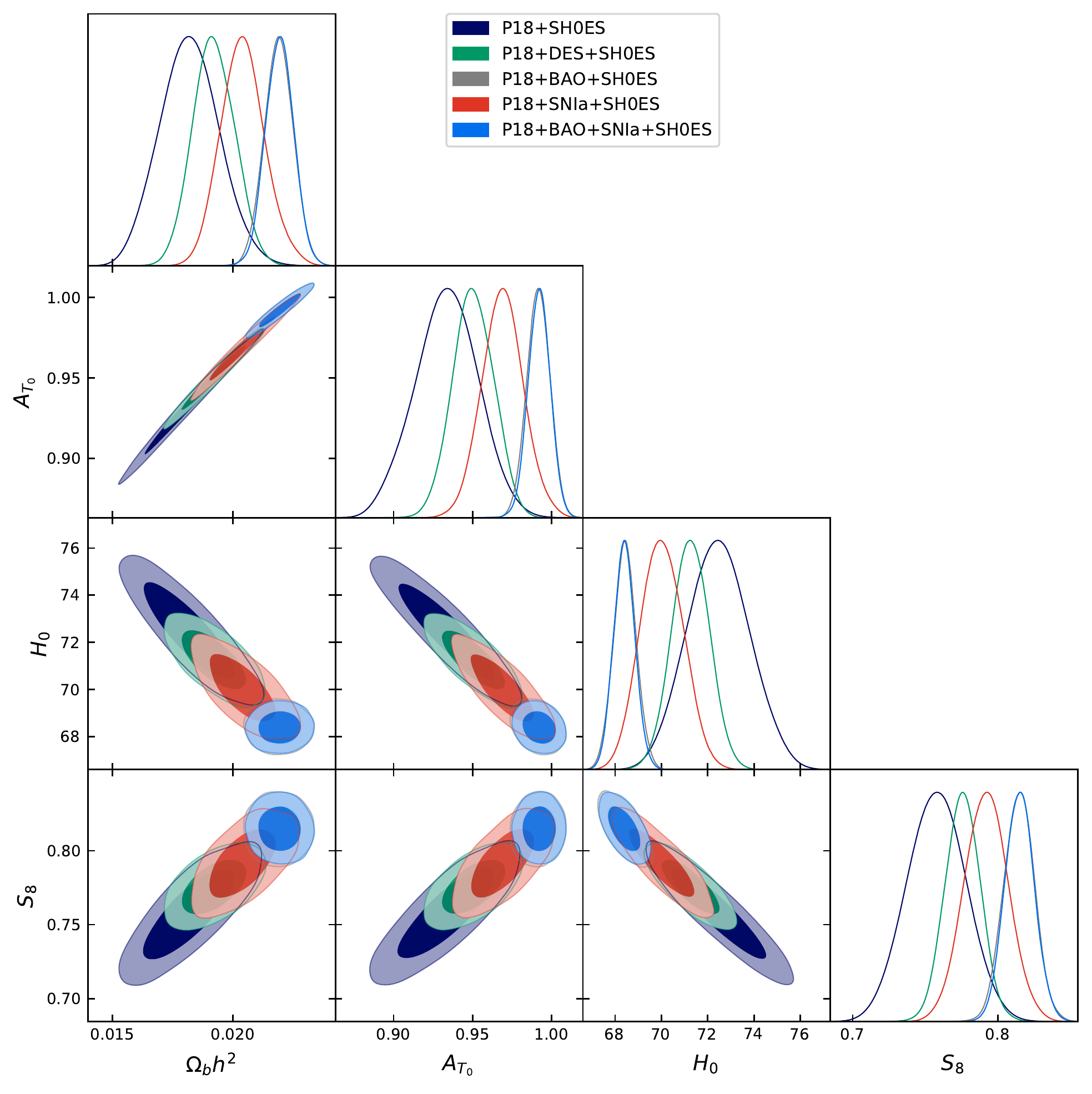}
\includegraphics[width=0.7\textwidth]{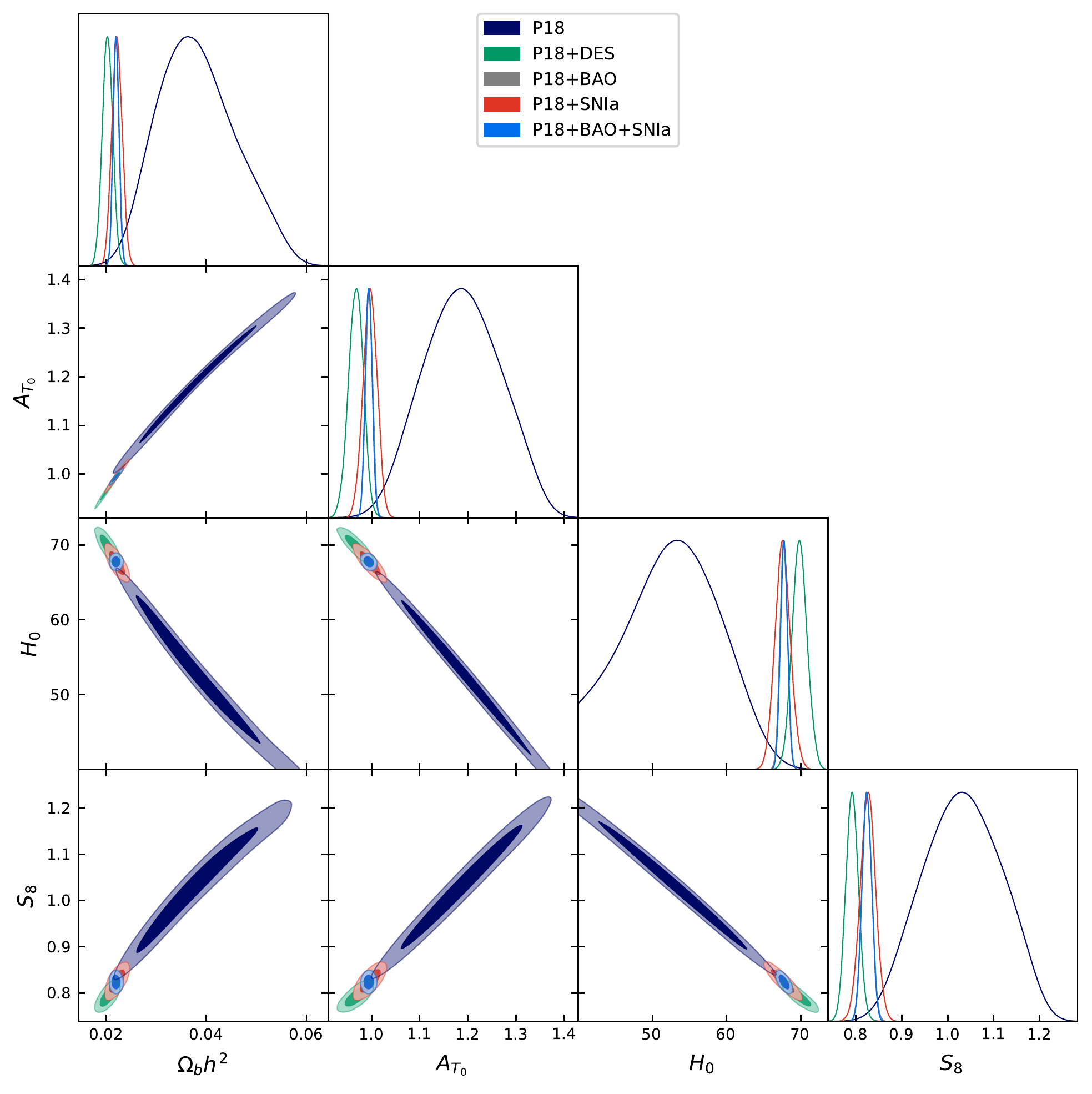}
\end{center}
\caption{Confidence contours and posterior distributions for selected 
parameters assuming the SM+T model: SH0ES prior included (top panel) or excluded 
(bottom panel).}
\end{figure}

\begin{figure}[h]
\begin{center}
\leavevmode
\includegraphics[width=0.7\textwidth]{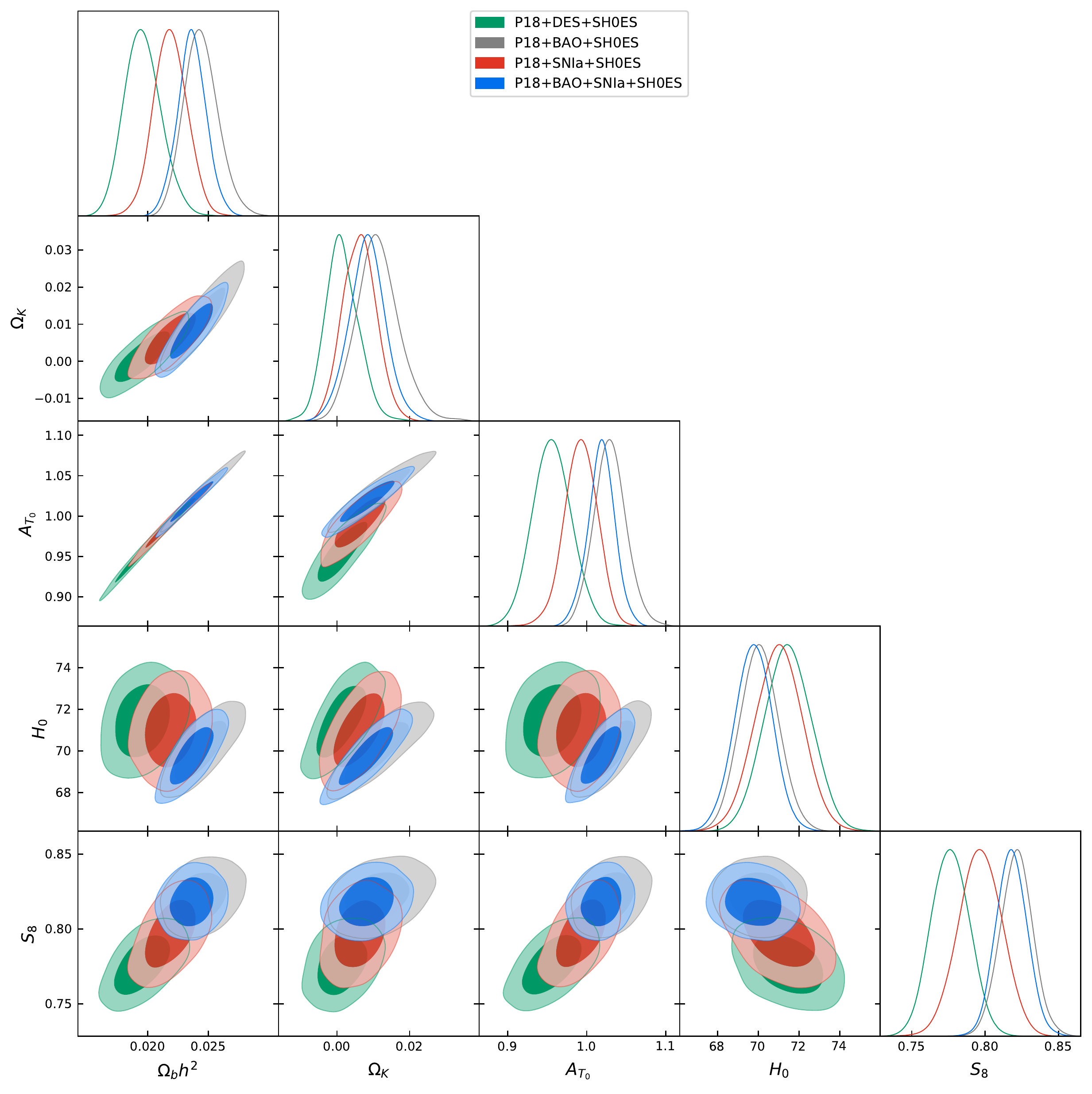}
\includegraphics[width=0.7\textwidth]{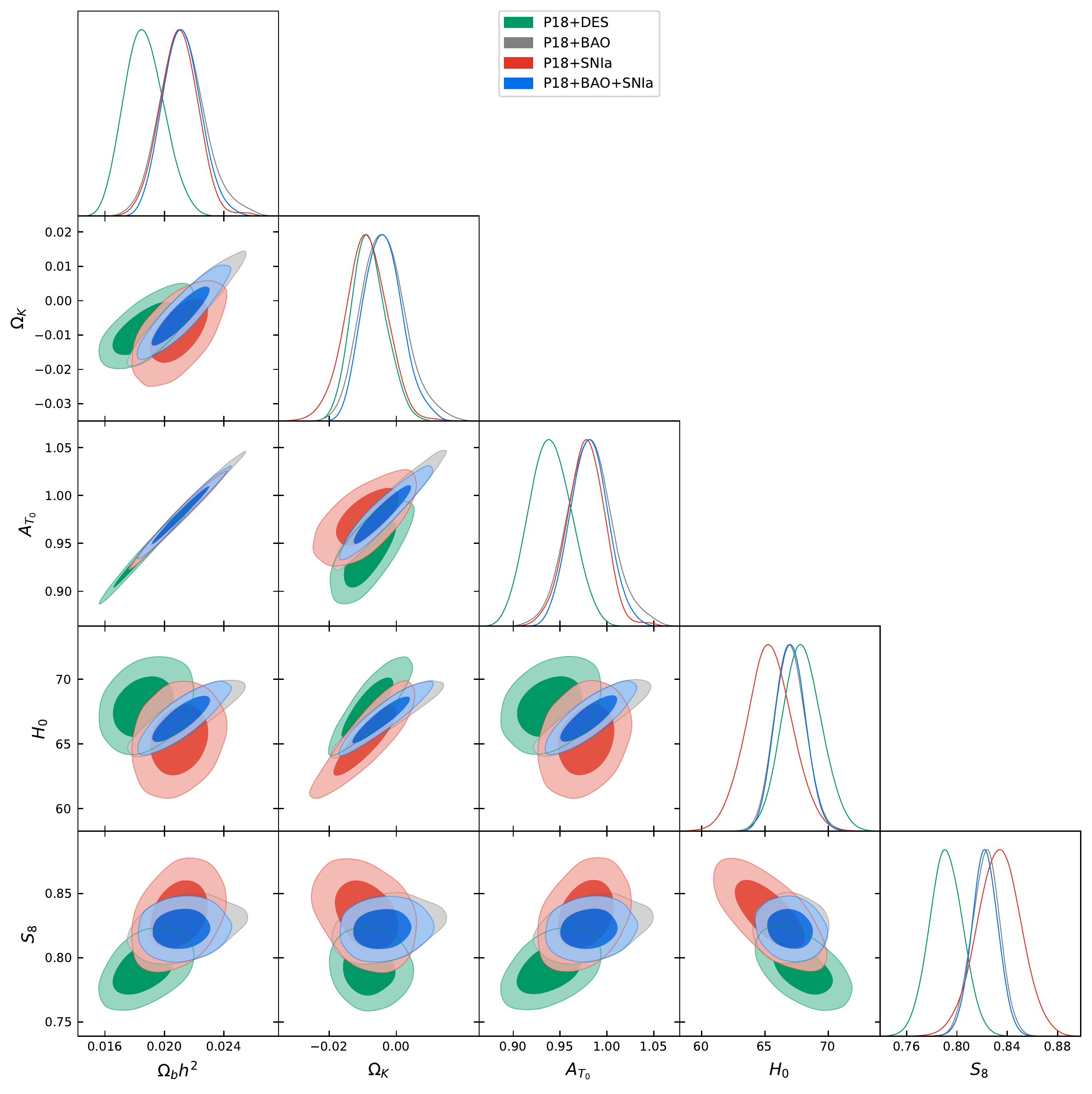}
\end{center}
\caption{Confidence contours and posterior distributions for selected 
parameters assuming the SM+K+T model: SH0ES prior included 
(top panel) or excluded (bottom panel).}
\end{figure}

\begin{figure}[h]
\begin{center}
\includegraphics[width=0.7\textwidth]{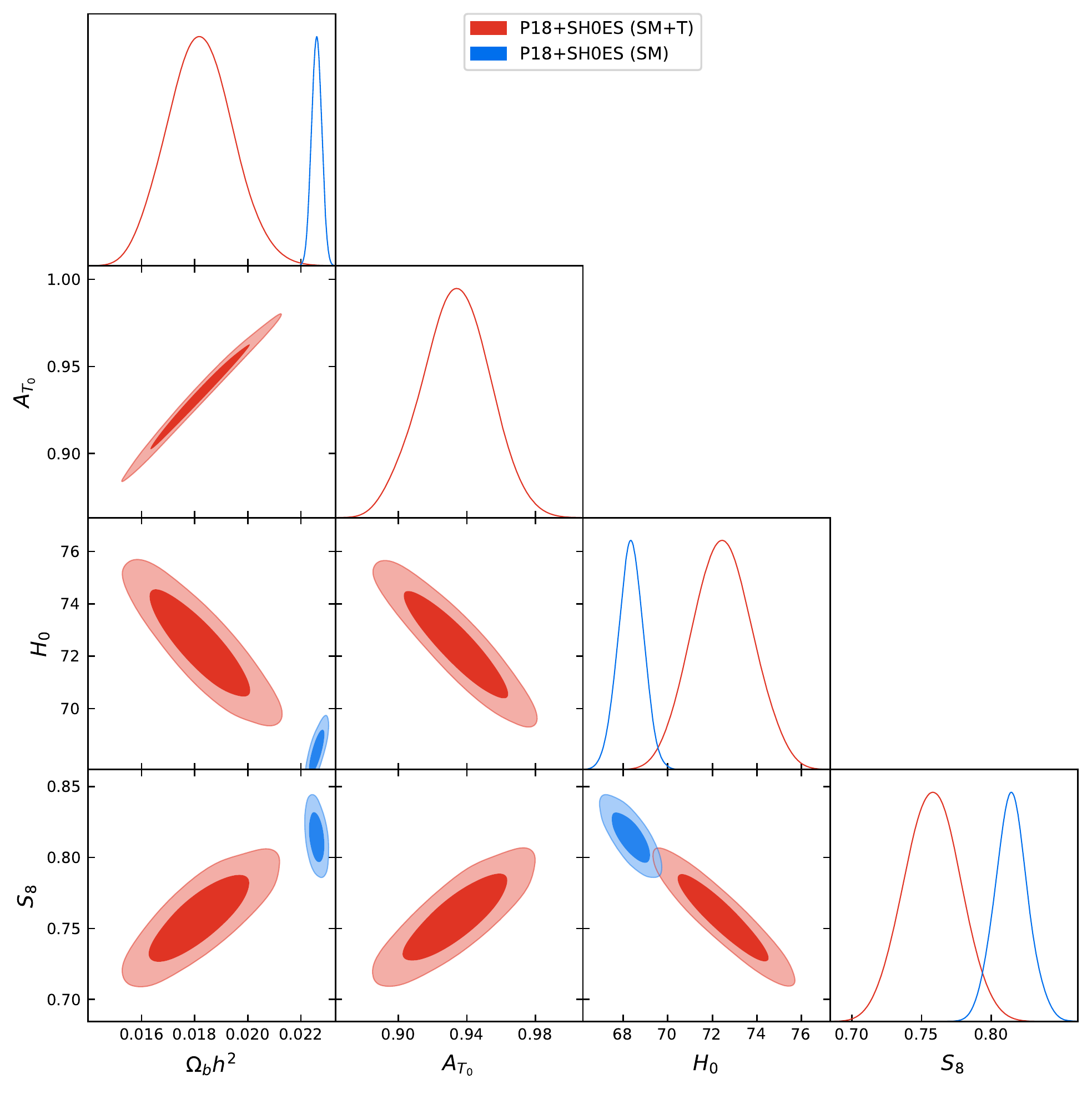}
\end{center}
\caption{Confidence contours and posterior distributions for selected 
parameters obtained for the SM+T and SM models extracted from 
the P18+SH0ES data.}
\end{figure}

Another statistical diagnostic that supports the SM+T (and to a smaller 
extent SM+K or SM+K+T) extension for data sets involving Planck \& DES 
is the value of $p_{D}$ (Eq. 4.2), the number of parameters actually 
constrained by the data set in question. For all data set combinations 
considered in this work, $p_{D}$ increases by 
$\lesssim 0.2-0.3$, except for the cases P18+DES \& P18+DES+SH0ES. 
In the former case $p_{D}$ increases 
by 0.23, 1.74 \& 1.04 when the SM is extended to SM+K, SM+T, and 
SM+K+T, respectively, while in the case of P18+DES+SH0ES it 
increases by 0.63, 0.96, and 1.07. From this perspective the 
addition of the temperature as a free parameter 
is warranted in both cases, especially when the SH0ES 
prior is excluded from the data. 
In comparison, extension of the SM by adding a free 
spatial curvature parameter only moderately increases $p_{D}$. 

The last five lines of Table III correspond to dataset combinations excluding the SH0ES 
prior. It is clear that in this case only the P18+DES combination 
benefits (equally well) from allowing for a free $A_{T_{0}}$ and 
non-flat geometry. The other three dataset combinations 
(P18+BAO, P18+SN, P18+BAO+SN) do not favor any of these SM extensions; 
the improved fit to the data is outweighed by the penalty 
(albeit small) incurred by adding either $\Omega_{k}$, 
or $A_{T_{0}}$, or both.  

Our interpretation of these results is as follows: P18+SH0ES favor 
a high value of $H_{0}$ which is anti-correlated with $A_{T_{0}}$, 
thus improving the fit when this parameter is added. With the DES 
dataset a systematically lower $S_{8}$ is obtained 
which is compatible with the other LSS datasets (e.g. [26]). 
This parameter is anti-correlated with $H_{0}$, which in turn anti-correlates with $A_{T_{0}}$. 
Therefore, the fit with either P18+DES+SH0ES and (to a lower extent) 
P18+DES is more statistically acceptable 
when $A_{T_{0}}\neq 1$. It is also clear from comparing $DIC_{SM+T}$ 
and $DIC_{SM+K}$ (Table III) when the SH0ES prior is imposed, that in 
those cases where the fit does not `benefit' from adding $A_{T_{0}}$, 
the fit improves somewhat by allowing for curved space, consistent with 
our expectation for a partial degeneracy between these two parameters. 
This is a consequence of the fact that a 
lower $T_{0}$ implies earlier recombination which could be partially mimicked 
by an open geometry ($\Omega_{k}>0$), where light rays are `defocused' and length 
scales look smaller than they really are. Thus, $A_{T_{0}}$ is expected to be correlated 
with $\Omega_{k}$; lowering $\Omega_{k}$ enhances focusing of the radiation, 
thereby compensating for the lower temperature effect. This strong correlation 
qualitatively explains the result of [11] that while the P18+SH0ES 
combination leads to a relatively low temperature,  
adding BAO data restores agreement with the local $T_{0}$ measurement. 
This is the case simply because BAO data result in a strong preference 
for flat space; fixing the temperature to its locally measured value 
results also in $H_{0}$ attaining the CMB-inferred value. Thus, it is 
clear that the parameter trio $A_{T_{0}}-H_{0}-\Omega_{k}$ is highly 
correlated, and that when the BAO dataset is added, all these parameters 
approach their `concordance' values.

It is important to determine 
whether there is an appreciable statistical preference (based on 
current datasets) for one of the extended models over the other. 
Our analysis shows that for virtually all data combinations 
(without the SH0ES constraint on $H_{0}$) adding the parameter $A_{T_{0}}$ 
to the SM results in a lower DIC than 
when $\Omega_{k}$ is added as a free parameter. This lends 
additional support to the idea that the small $\Omega_{k}$-tension 
between Planck and, e.g. BAO data, is no more than a tension between 
local inference of $T_{0}$ versus the value inferred from CMB 
features imprinted at recombination, i.e. that the interpretation 
of closed spatial geometry advocated in [25] could be replaced 
by the statement that, assuming that space is flat, $T$ 
at recombination is higher than is thought based on adiabatic 
evolution history. Nevertheless, our analysis shows that except 
for the P18+DES, data combinations that exclude the SH0ES constraint 
clearly indicate no preference for both SM extensions as they result 
in even slightly higher DIC values. More interesting in that respect 
are dataset combinations that do involve the SH0ES 
constraint. In this case, either P18+SH0ES or P18+DES+SH0ES results 
in a better fit with SM+T, whereas P18+BAO+SH0ES, P18+SN+SH0ES, 
or P18+BAO+SN+SH0ES yield better fits to the SM+K+T model. Thus, when 
the SH0ES constraint is included there is no clear 
preference for either model extension.

Figure 1 is a triangle plot that shows the 1-\& 2-$\sigma$ 
contours and posterior distributions of several 
parameters for the SM+T model. Shown are results for the P18, 
P18+DES, P18+BAO, P18+SNIa and P18+BAO+SNIa dataset combinations 
with (upper panel) and without (lower panel) SH0ES prior. 
Corresponding results (except for the case of Planck alone) for 
the SM+K+T model are shown in Figure 2. Allowing for a free 
$A_{T_{0}}$ results in a factor of a few larger uncertainties on 
several parameters, e.g. $H_{0}$. In addition, certain 
parameters correlated with $A_{T_{0}}$ are in a $\sim 2-3\sigma$ 
tension with those obtained from the SM-based analysis with the 
locally determined $T_{0}$. 

As we have shown, $H_{0}$ is strongly correlated with $A_{T_{0}}$; 
according to our analysis using the recent P18+SH0ES data 
$H_{0}$ increases to $\gtrsim 72.5\pm 1.3$ km/sec/Mpc, as compared 
to the value deduced for the `vanilla' SM ($\Omega_{k}=0$), 
$68.34\pm 0.57$ km/sec/Mpc. 
Allowing for non-vanishing spatial curvature only slightly changes this value from 
$71.3\pm 1.2$ to $72.3 \pm 1.4$ km/sec/Mpc. Therefore, while allowing for a 
non-vanishing curvature significantly alleviates the Hubble tension, this cannot be 
equally said for the CMB temperature. Whereas the correlation with $A_{T_{0}}$ 
brings the cosmologically-inferred $H_{0}$ closer to its locally-inferred value, 
perhaps more important is the fact that the uncertainty in 
this parameter increases by a factor $\sim 3$ compared to the value deduced 
in the standard analysis. These two changes weaken the Hubble tension from $\gtrsim 5\sigma$ to only 
$\sim 1-2\sigma$ with the local inference of $H_{0}$ from strong lensing and SH0ES, and to 
$\lesssim 1\sigma$ with other local probes. The degeneracy of $A_{T_{0}}$ with $S_{8}$ 
and $H_{0}$ brings them closer to locally measured values, thereby alleviating 
long-standing tensions of cosmologically versus locally inferred values.  

The impact of adding either $\Omega_{k}$ or $A_{T_{0}}$ to the best-fit SM 
can be clearly illustrated by considering 
$\chi_{SH0ES}^{2}$, which is calculated by CosmoMC when the SH0ES prior is 
imposed. In the case of P18+SH0ES data combination we obtain 
$\chi_{SH0ES}^{2}=16.2\pm 3.2$, $4.3\pm 3.5$ and $2.1\pm 2.3$ 
for the SM, SM+K and SM+T models, respectively. A qualitatively similar impact 
is seen for the P18+DES+SH0ES combination, for which 
the corresponding values are $13.5\pm 2.6$, $2.8\pm 2.3$, $4.1\pm 2.3$, and $3.9\pm 3.0$, 
respectively. Assuming a-priori the fixed values $\Omega_{k}=0$ and $A_{T_{0}}=1$ 
yields a low $H_{0}$, and consistency with the SH0ES prior results in a $\chi^{2}$ 
value that is higher by $O(10)$. When these parameters are allowed to vary, a significantly 
lower $\chi^{2}$ and (even more relevant) overall DIC values are obtained.

\section{Summary}

Despite the remarkable success of the SM, over the last decade significant evidence has 
mounted that various degrees of tension exist between locally inferred cosmological 
parameters and inference from the earlier universe. While it is clearly of interest to 
explore the possibility that these tensions could be at least partially 
resolved, our other key objective is the inference of $T(z)$ 
at $z\approx 1100$ as part our ongoing interest in direct determination of $T(z)$ 
over the widest possible redshift range.  The analysis reported here 
has resulted in an improved determination of the CMB temperature at recombination 
in both flat or curved space, with and without the local SH0ES prior on $H_{0}$.

Whereas the physics of recombination is very sensitive 
to the CMB temperature correlations of $A_{T_{0}}$
with most of the other key cosmological parameters limits the precision of 
$A_{T_{0}}$ extraction from the P18 dataset 
to the percent level. In principle, the spectrum of the thermal component of the 
SZ effect could be an ideal probe of the evolution of the 
temperature with redshift due to its exponential dependence on the temperature 
and lack of correlation with other cosmological parameters. However, high-quality 
measurements of the effect towards clusters at 
sufficiently high redshifts are rare and 
certainly not competitive with the $z_{*}\approx 1100$ leverage 
provided by CMB anisotropy. In addition, marginalization over 
the cluster comptonization parameter and gas temperature 
(the latter is required for marginalization over relativistic corrections to the 
non-relativistic effect), 
as well as over the cluster bulk velocity, significantly degrade 
the nominal achievable precision based on the spectral shape of 
the thermal effect. Nevertheless, near future surveys with e.g. the Simons Observatory 
are expected to detect thousands of clusters; this 
opens up the possibility of complementing CMB anisotropy 
with an independent probe that continuously 
samples T(z) up to $z\approx 2$. 
Of course, the efficiency of this probe is expected 
to depend on the specific non-standard T(z) models.

Independent 
inference of $T$ at recombination is an important consistency test of the SM, and of 
the assumption of adiabatic evolution history in particular. Unlike 
the case of local measurements that are based on 
the steep dependence of the Planck spectrum on $T$, 
the inference from temperature anisotropy and polarization at $z\approx 1100$ 
carried out in this and similar recent works relies only on achromatic angular 
power spectra aided by luminosity distance measurements, as well as other 
observables that do not explicitly depend on the CMB temperature. 
In addition, the temperature perturbation measurements, from which we extract 
$T(z_{*})$, are $O(10^{-5})$ times smaller than $T_{0}$. Nevertheless, the 
$68\%$ uncertainty level on the inferred temperature is only $O(10^{2})$ larger 
than the uncertainty obtained from local measurements due to the strong dependence 
of the CMB power spectra on $T(z_{*})$. This $\sim 1\%$ precision elevates 
precision determination of $T(z_{*})$ to the current typical precision level of other 
cosmological parameters.

Insight gained on the Hubble tension is that treating the CMB temperature as a 
free parameter considerably lowers the previous $\sim 5\sigma$ tension at a 
`cost' of a moderately lower temperature, at the $2\sigma$ level, than would be 
expected if the locally measured $T$, extrapolated to $z\approx 1100$, holds 
at the FIRAS-deduced precision level. Similarly reduced level of Hubble tension 
can also be achieved by a slightly positive value of $\Omega_{k}$, i.e., in a model 
with geometrically open spacetime. Without the local Hubble prior all dataset 
combinations considered in this work favor either a lower $T$ at last scattering, or the FIRAS value, 
and either flat or closed universe. Notable departures from the `vanilla' values are 
obtained for SM+T model when contrasted with the P18 \& P18+DES datasets; 
in this case $A_{T_{0}}=1.189\pm 0.080$ \& $A_{T_{0}}=0.968\pm 0.015$, respectively, 
where the former nicely illustrates the $\Omega_{k}-T$ 
correlation. For the SM+K mode with only Planck data are included, $A_{T_{0}}=1$, but 
only if space is flat. The $\gtrsim 2\sigma$ higher CMB 
temperature at recombination that we find here will increase to $\sim 3\sigma$, if 
$\Omega_{k}$ increases in a SM+K+T model (as found also in [12]). Much like in the SM+K model, 
we obtain in the SM+T model the anomalously small value $H_{0}=53\pm 6$ km/sec/Mpc.

Allowing for non-vanishing curvature in the SM+K+T model we 
obtain $\Omega_{k}=-0.0081_{-0.0056}^{+0.0046}$ \& $A_{T_{0}}=0.940\pm 0.022$ 
when the P18+DES dataset combination. 
When the same model is contrasted with the P18+SN dataset 
combination we deduce $\Omega_{k}=-0.0089\pm 0.0063$ \& $A_{T_{0}}=0.977\pm 0.020$. 
In these cases $A_{T_{0}}$ is $\sim 1\sigma$ lower than in the corresponding dataset 
combinations in the SM+T model, simply due to the $\Omega_{k}-T$ correlation; imposing 
flat space artificially increases the temperature at last scattering. 
In the same vein, these same dataset combinations are consistent with flat space if the SM+K model is considered, again 
due to the $\Omega_{k}-T$ correlation; letting the temperature to be different than the FIRAS value 
allows both $\Omega_{k}$ \& $T$ to drop to their data-favored values. 

The above result of a possibly higher temperature at recombination than extrapolated 
from local measurements (under the assumption of adiabatic evolution) has an intriguing 
implication: The observed locations of the acoustic peaks in the anisotropy power spectrum 
can only be consistent with a higher temperature if (global) 
spatial curvature is negative, i.e. if the the radiation is de-focused. If so, then by not 
allowing the temperature to be a free parameter would result in a larger curvature radius, 
infinite (flat space), or even a closed universe. The preference seen in 
the Planck data for 
a spatially closed geometry is based on this exact premise -- that the comoving temperature 
at recombination is fixed to its locally inferred value. Therefore, it is possible that the 
deduced `closed' space at the $\sim 2\sigma$ confidence level actually implies that space 
is flat, but that $T$ at recombination is $\sim 2\sigma$ (i.e. a few percent) higher than 
inferred locally. Since other cosmological probes are sensitive to the spatial geometry, 
but are insensitive to $A_{T_{0}}$, they all favor flat space (in clear contrast with the 
indication from the Planck data). We conclude that it is a viable possibility that -- rather 
than suggesting a closed spatial geometry -- the Planck result could be interpreted as 
an indication for a non-standard evolution of the CMB temperature, i.e. that the universe has 
been cooling down somewhat faster than expected in the SM.

\section*{Acknowledgments}
We gratefully acknowledge Dr. Sharon Sadeh for his indispensable expert help 
in installing and running CosmoMC.
MS would like to thank Dr. Nissan Itzhaki for constructive discussions 
when this work was being contemplated.
This research has been supported by a grant from the Joan and 
Irwin Jacobs donor-advised fund at the JCF (San Diego, CA).

\end{document}